\documentclass{icbo}

\usepackage{xspace}

\usepackage{listings}
\usepackage{tawny}
\lstMakeShortInline[style=tawnystyle]|
\lstnewenvironment{tcode}[1][]%
{\lstset{style=tawnystyle,frame=shadowbox,basicstyle=\footnotesize\ttfamily,#1}}{}

\newcommand{\tawny}{Tawny-OWL\xspace}
\newcommand{\protege}{Prot\'eg\'e\xspace}

\usepackage{hyperref}

\begin{document}
\title[Scaffolding the Mitochondrial Disease Ontology]{Scaffolding the
  Mitochondrial Disease Ontology from extant knowledge sources}

\author[J. D. Warrender and P. Lord]{Jennifer D. Warrender and Phillip
  Lord\footnote{To whom correspondence should be addressed:
    phillip.lord@newcastle.ac.uk}}

\address{School of Computing Science, Newcastle University,
  Newcastle-upon-Tyne, UK}

\maketitle

\begin{abstract}
  Bio-medical ontologies can contain a large number of concepts. Often
  many of these concepts are very similar to each other, and similar
  or identical to concepts found in other bio-medical databases. This
  presents both a challenge and opportunity: maintaining many similar
  concepts is tedious and fastidious work, which could be
  substantially reduced if the data could be derived from pre-existing
  knowledge sources. In this paper, we describe how we have achieved
  this for an ontology of the mitochondria using our novel ontology
  development environment, the \tawny library.
\end{abstract}

\section{Introduction}

Bio-medical ontologies vary in size, with largest containing millions
of concepts. Building ontologies of this size is complex,
time-consuming and expensive and just as challenging to maintain and
update.

Ontologies are only one of many mechanisms for the computational
representation of knowledge. In some cases, ontologies are created
where many of the needed concepts will be available elsewhere as terms
in different structured representations. Being able to reuse these
representations as a \textit{scaffold} for the rest of an ontology
might be able to reduce the cost and work-load of producing
ontologies.

This is evidenced by, for instance, SIO~\citep{sio} which contains a
list of all the chemical elements. Or the Gene Ontology
(GO)~\citep{go}, which contains many terms related to chemical
homeostasis, each of which need to relate to a specific chemical
described in ChEBI~\citep{chebi}. In addition to being described
elsewhere, these concepts are often highly similar to each other. In
extreme cases such as the amino acid ontology~\citep{greycite9379},
ontologies can consist of only related concepts, and ``support''
concepts that are used to describe them.

One solution to this is the use of patterns. A pattern is an abstract
specification of an ontology axiomatisation with a number of
``variables''. The pattern is instantiated by providing values for
these variables, which are then expanded into the full axiomatisation
providing one or more concepts.

Patterns have been implemented by a number of different tools, which
differ in how the patterns are specified, and how and when the values
are provided for the variables. For example, \textit{termgenie} is a
website which allows submission to GO (and
others)~\citep{Dietze_2014}. Variable values are entered through a
form which then generates axioms, definitions and
cross-references. For instance, this is the axiomatisation from
termgenie when defining the term ``cytosine homeostasis''

\begin{verbatim}
is_a: GO:0048878 {is_inferred="true"} 
  ! chemical homeostasis
intersection_of: GO:0048878
  ! chemical homeostasis
intersection_of: 
  regulates_levels_of CHEBI:16040 ! cytosine
relationship: 
  regulates_levels_of CHEBI:16040 
  {is_inferred="true"} ! cytosine 
\end{verbatim}

As well as the axiomatisation, termgenie also generates a number of
different annotations including a definition, submitter information,
and status. With termgenie, patterns are specified through the use of
JavaScript functions.

In addition to termgenie, other systems also allow patterns. For
example, both the desktop and web version of \protege contain forms,
which grant users the ability to customise the GUI and specify several
axioms at once. In this case, patterns are declaratively defined
(implicitly, with a GUI design) in
XML~\citep{tudorach_icd_webprotege}. Applications like
Populous~\citep{Jupp_Wolstencroft_Stevens_2011} and
Rightfield~\citep{rightfield} use spreadsheets or spreadsheet-like
interfaces to enter data, which is then transformed into a set of OWL
axioms based on a pattern. In the case of these two, the patterns are
specified in OPPL, a pattern language for OWL which can also be used
independently~\citep{aranguren_Stevens_Antezana_2009}. Finally, the
Brain API allows programmatic construction of ontologies in an easy to
use manner using Java~\citep{croset2013}.

While these systems are all aimed at somewhat different use-cases,
they all address the same problem; how to produce a large number of
concepts all of which are similar, and to do so with a high-degree of
repeatability. However, the use of this form of patternised ontology
tool presents a number of problems. These tools provide a mechanism
for adding many axioms at once, but not removing them
again\footnote{OPPL can remove axioms as well as add them but this is
  not automatic.}. If the knowledge changes, then this is a problem as
the axioms added from a given pattern need to be removed or
updated. Furthermore, if the knowledge engineering changes i.e. the
pattern is updated, then all axioms added from any use of the pattern
must also be updated.

In this paper, we describe how we have addressed these problems with the
Mitochondrial Disease Ontology (MDO), through the use of the \tawny
environment, which is a fully programmatic environment for ontology
development. With \tawny, we can use a \textit{pattern-first} ontology
development process, building with patterns and data from extant knowledge
sources from the start. This has allowed us to generate a
  \textit{scaffold} which we can then populate further with hand-crafted links
  between parts of this scaffold where the knowledge exists. As a result, it is
possible to update both the knowledge and the patterns by simply regenerating
the ontology. This process promises to aid in both the construction and
maintenance of ontologies.

\begin{sloppypar}
The MDO is available from
\url{https://github.com/jaydchan/tawny-mitochondria}. \tawny is
available from \url{https://github.com/phillord/tawny-owl}.
\end{sloppypar}

\section{The Mitochondria Disease Ontology (MDO)}
\label{sec:mitoch-dise-ontol}

Mitochondria are complex organelles found in most eukaryotic
cells. Their key function is to enable the production of ATP through
oxidative phosphorylation, providing usable energy for the rest of the
cell. The mitochondria carry their own small genome containing 37
genes in human. Many other genes are involved in producing proteins
involved in mitochondrial function, but these are encoded in the
nuclear genome. A number of mitochondrial genes are associated with
diseases; the first identified of these is the MELAS~\citep{melas},
which is most commonly caused by a point mutation in a tRNA found in
the mitochondrial genome.

As with many areas of biology, mitochondrial research is a large,
knowledge-rich discipline. Our purpose with the MDO is to attempt to
formalise this knowledge, using an incremental or ``pay-as-you-go''
data integration approach. The ontology here serves as a tool for
reasoning and knowledge exploration, rather than to form as a
reference ontology~\citep{handbook2}.  This is an approach we have
previously found useful in classifying
phosphatases~\citep{wolstencroftetal2006}. The hope is that we can
incorporate new knowledge as it is released, checking it for
consistency and cross-linking it with existing knowledge.

\section{\tawny}
\label{sec:tawny}

In this section, we give a brief description of \tawny~\citep{tawny}
and how it supports pattern-first development. \tawny is a library
written in Clojure, a dialect of lisp. It wraps the OWL
API~\citep{owlapi} and allows the fully programmatic constructions of
ontologies. It has a simple syntax which was modelled on the
Manchester Syntax~\citep{ms2}, modified to integrate well with
Clojure. It can be used to make simple statements in OWL:

\begin{tcode}
(defclass A :super (some r B))
\end{tcode}

which makes defines a new class $A$ such that
$A\sqsubseteq~\exists~r~B$. Although this is similar to the equivalent
Manchester Syntax statements, \tawny provides a feature called
``broadcasting'' which is, essentially a form of pattern. So this
following statement:

\begin{tcode}
(some r B C)
\end{tcode}

is equivalent to the two statements $\exists~r~B$ and
$\exists~r~C$. We apply the first two arguments (|some| and |r|) to
the remaining ones consecutively.  It also provides simple patterns,
such as the covering axiom, so:

\begin{tcode}
(some-only r B C)
\end{tcode}

is equivalent to three statements $\exists~r~B$, $\exists~r~C$ and
$\forall~r~(B~\sqcup~C)$. While the patterns shown here are provided
by \tawny, end ontology developers are using the same programmatic
environment.  Patterns are encoded as functions and instantiated with
function calls. For instance, we could define |some-only| as follows:

\begin{tcode}
(defn some-only [property & classes]
 (list (some property classes)
       (only property
             (or classes))))
\end{tcode}
           
Here |defn| introduces a new function, |property & classes| are the
arguments, and |list| packages the return values as a list. |some|,
|only| and |or| are defined by \tawny as the appropriate OWL class
constructors.

It is, therefore, possible to build \textit{localised patterns} ---
custom patterns for use predominately with the current
ontology~\citep{warrender_thesis_2015}. Patterns can call each other
and can be of arbitrary complexity. The use of \tawny, therefore,
inverts the usual style of ontology development. Non-patternised
classes are just trivial instantiations of patterns.

\section{Building a Mitochondrial Scaffold}
\label{sec:build-mitoch-scaff}

Following a requirements gathering phase for MDO, it was clear from
our competency questions (for example ``What are all the genes/proteins
that are associated with a specific syndrome?'') that we needed many
concepts which were heavily repetitive, and further which have
comprehensive and curated lists available. We describe these parts of
the domain knowledge as the \textit{scaffold}. For example, there are
around 761 genes whose products are involved in mitochondrial
function. Classes representing these genes do not, in the first
instance, require complex descriptions, and are defined within MDO as
follows:

\begin{tcode}
(defclass Gene)

(defn gene-class [name]
 (owl-class name :label name :super Gene))
\end{tcode}

This pattern is then populated using a simple text file, with the 761
gene names present. The gene pattern is an extremely simple pattern,
as these concepts are self-standing. Other parts of the ontology are
even simpler; for instance, for describing mitochondrial anatomy, the
classes have similar complexity to the genes, but there are only
15. In this case, classes are defined with a pattern and a list
``hard-coded'' into the MDO source code, rather than using an external
text file. Other patterns are more complex. For instance, the
subclasses of |Disease| are defined as follows:


\begin{tcode}
(defn disease-class [name omim lname]
 (let [disease
       (owl-class name
                  :label name
                  :super Disease)]
  (if-not (nil? omim)
   (refine disease
           :annotation 
           (see-also 
            (str "OMIMID:" omim))))
  (if-not (nil? lname)
   (refine disease
           :label 
           (str "Long name:" lname)))))
\end{tcode}
               
This function adds two annotations to each disease class, if they are
available. This function also demonstrates the use of conditionals
(|if|), predicates (|nil?|) and string concatenation (|str|); these
are not provided by \tawny, but by Clojure and demonstrate the value
of building \tawny inside a fully programmatic environment.

\section{Fitting out the Scaffold}
\label{sec:what-we-have}

\begin{figure}[]
  \centering
  \includegraphics[width=\columnwidth,keepaspectratio=true]{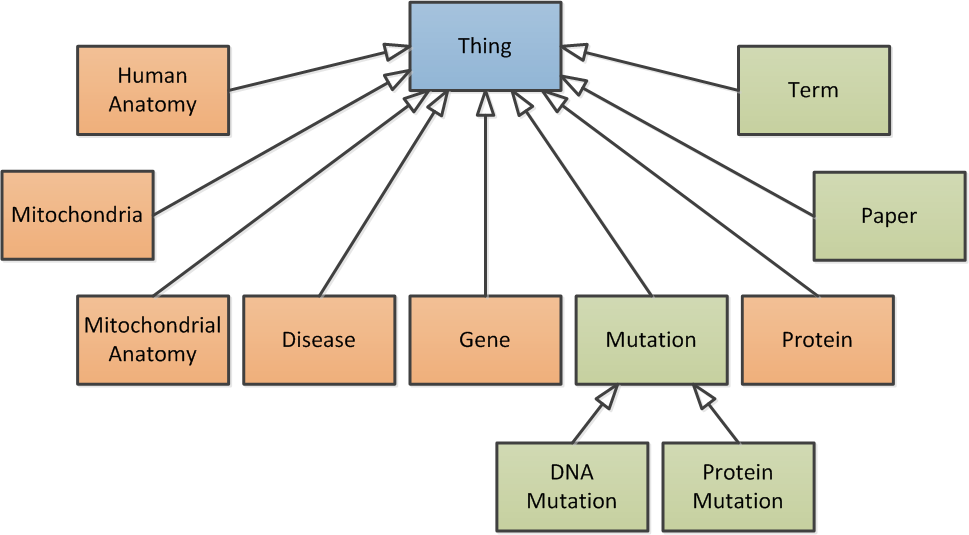}
  \caption[Top-level structure]{The top-level structure of Mitochondrial
    Disease Ontology. Classes that are a part of the scaffold are
    coloured in orange, while classes that are built on top of the
    scaffold are coloured in green.}
  \label{fig:mtoplevel}
\end{figure}

The top-level of the MDO is shown in Figure~\ref{fig:mtoplevel}. Of these
classes, ``Paper'' and ``Term'' are described later.

The remaining classes define the scaffold, which now has a total of 1357
classes; a break-down of these classes and their sources is shown in
Table~\ref{tab:genericStats}.

\begin{table}[h]
  \processtable{Table showing the type, number of and data source for each generic mitochondrial ontology class\label{tab:genericStats}}
  {\begin{tabular}{lll}
     Class type & Count & Data source \\
     \noalign{\smallskip} \hline \noalign{\smallskip}
     Disease & 41 & \href{http://www.umdf.org/site/c.8qKOJ0MvF7LUG/b.7934629/k.4C9B/Types_of_Mitochondrial_Disease.htm}{UMDF website} \\
     Gene & 761 & \href{http://www.ncbi.nlm.nih.gov/gene}{The NCBI Gene portal} \\
     Human Anatomy & 61 & \href{http://www.unifr.ch/ifaa/}{The Terminologia Anatomica.} \\
     Mitochondrial Anatomy & 15 & \href{http://www.newcastle-mitochondria.com/mitochondria/what-do-mitochondria-do/}{Mitochondrial Research Group} \\
     Protein & 479 & \href{http://www.uniprot.org/}{UniProt} \\
 \end{tabular}}
\end{table}

For the next stage of the process, we are now building on top of this
scaffold, using hand-crafted and bespoke knowledge. This is being
achieved by manual extraction of knowledge from papers about
mitochondria. Our initial process is to find references in papers to
the terms that are represented by classes we have built in the
scaffold, and draw explicit relationships between these papers and the
scaffolded knowledge that they describe. Currently, these classes also
use a patternised approach; the raw data is held in a bespoke (but
human readable) syntax\footnote{In this case \textit{EDN} which is a
  text representation of Clojure data structures; it looks rather like
  JSON.}, which is then parsed and used to instantiate patterns. In
total, there are now 2174 classes created from this approach from
around 30 papers. These terms currently are not defined beyond their
name and the source paper from which they were identified. We do not
consider them directly as part of the scaffold, as they are not from
an extant knowledge source, but one that we have created; they are the
first layer build on top of our scaffold. We expect future layers to
use the \tawny syntax directly, as the knowledge increases in
complexity and decreases in regularity.

\section{Resiliance to Change}
\label{sec:resiliance-change}

One key feature of our development process is that the OWL which
defines the MDO is no longer \textit{source code} but
generated. Rather it is generated from patterns defined in \tawny and
text files which are used to instantiate these patterns. The in-memory
OWL classes and associated OWL files are generated on-demand, by
\textit{evaluating} the patterns. Effectively, we regenerate the
ontology every time we restart the environment. In this section, we
consider the types of changes that can happen, and how these changes
impact on MDO.

The scaffold of MDO is sensitive to changes in its dependency
knowledge sources. First, new terms can be entered into extant
sources, which will necessitate the addition of new classes. For the
MDO, this simply necessitates re-importing the knowledge. The addition
of equivalent new classes will then happen automatically according to
the patterns already defined; no other changes should be necessary for
the MDO, although we may wish to refer to the new classes in other
parts of the ontology.

Second, terms may be removed from dependencies; so, for example, a disease may
be redefined by the UMDF. In many cases, for the MDO, this is not problematic
-- the equivalent classes will simply disappear from the ontology. \tawny
provides two features to help with changes to terms in the scaffold when these
terms are also referred to outside of the scaffold. \tawny uses a
``declare-before-use'' semantics, so removal of classes from the scaffold will
cause fail-fast behaviour when they are used elsewhere. The Brain environment
uses the same semantics for similar reasons~\citep{croset2013}. In addition,
\tawny provides a ``deprecation'' facility which allows the developer to
continue refer to terms from the scaffold which have been removed, but to
receive warnings about this use; this is rather like obsolescence, but happens
automatically\footnote{\tawny is implemented in a Lisp and so is homoiconic;
  this makes it particularly straight-forward to automate code updates if we
  choose.}.

Third, the MDO scaffold can also cope straight-forwardly with changes
to patterns.  As with the addition or removal of terms from
dependencies, pattern changes will simply take place by re-evaluating
the ontology.

Finally, the MDO is resilient to changes in ontology engineering
conventions.  For example, MDO does not use OBO style numeric
identifiers, nor provide stable IRIs for integration with linked data
sources since these are not critical at the current time\footnote{Our
  initial intention was to use PURLS from \url{www.purl.org} but have
  found practical problems with generating these.}. They, however,
could be added easily to all existing (and future) terms in a few
lines of code, using an existing facility within \tawny for minting
and persisting numeric identifiers in an automatic, yet managed,
way. This change would just alter IRIs and would have no impact on
references between concepts inside or outside of the scaffold.

In conclusion, as well as enabling rapid construction of the MDO, we
believe that the pattern-first scaffolding approach should also allow
easy maintenance of the ontology.

\section{Discussion}
\label{sec:discussion}

In this paper, we have described how we have used a number of extant
knowledge sources, combined with patterns defined using the \tawny
library to rapidly, reliably and repeatedly construct a scaffold for MDO.

We have previously used a related patternised methodology to construct
a complex ontology describing human chromosome rearrangements
(i.e. The Karyotype Ontology
(KO)~\citep{warrender-karyotype}). However, unlike KO, the
mitochondrial knowledge we want to encapsulate is found in numerous
independent sources (e.g. published papers and online databases) and
in a variety of formats (e.g. ``free text'' and CSV); the use of
several patterns to form a scaffold is unique to MDO. Conversely, the
axiomatisation of MDO from these sources is simple; this cannot be
said for KO, most of which is generated from a single large
pattern~\citep{warrender-pattern}. In addition, while our knowledge of
the karyotype is constrained and is essentially finished, the
community's understanding of mitochondria and mitochondrial disease is
incomplete and will grow in response to the demands of changing
knowledge.

This methodology is extremely attractive for a number of
reasons. First of all, it allows a very rapid way of scaffolding an
ontology for a complex area of knowledge. At this stage, most of the
classes created are simple and self-standing, although in some cases
do have relationships to other entities in the scaffold. At this
point, we have built the ontological equivalent of a data warehouse:
terms have been taken from elsewhere and have undergone a form of
schema reconciliation into ontological classes.

One key feature of the MDO is that it has been built using tools
designed for software development; these tools are relatively advanced
and well-maintained\footnote{And, usefully, not dependent on academic
  developers for future maintenance.}~\citep{tawny}. Moreover,
recreating the MDO ontology from our original \tawny source code is an
intrinsic part of the development process; there is no complex release
process and any ontology developer can recreate the OWL file with a
single command. While, the system as it stands has a high-degree of
replicability, the design decisions implicit in the source code are
not necessarily apparent. For the basic scaffold this is, perhaps, not
a major issue, however as MDO is developed outside of its scaffold ,
we expect to integrate more documentation into the source code itself,
using \textit{lentic}, a recently developed tool for literate
programming~\citep{greycite23590}.

We believe that the engineering process that we have used to build the
scaffold is resilient to change, as described in
Section~\ref{sec:resiliance-change}. Despite this resilience, our use
of external sources of knowledge does bring with it new dependencies,
with all the issues that this entails for change management. We
believe that we can manage this by borrowing best practice from
software engineering. Importing knowledge into the scaffold can, in
many cases, happens entirely automatically from our extant knowledge
sources. Considering just the gene lists, we can either import from a
local, fixed copy of this list, or take the current version live from
the NCBI portal. In software engineering terms, the former is a
\textit{release dependency} and provides stability, while the latter
is a \textit{snapshot dependency} which will fail-fast, allowing rapid
incorporation of new knowledge. The latter is particularly useful
within a continuous integration environment which are used with other
ontologies~\citep{greycite2899}, and are also fully supported
by~\tawny\citep{tawny}.

Although we have not described its usage here, with the MDO we are not
forced to use \tawny for all development. It would be possible to
combine predominately hand-crafted development using \protege, for
instance, with some patternised classes; for example, the OBI uses
this approach~\citep{2041-1480-1-s1-s7}. For, the MDO, in fact almost
all terms other than the top-level has been created from other
syntaxes, generally a flat-file. For larger projects, we envisage that
most ontology developers would not need to use the programmatic nature
of \tawny. While we appreciate the value of a single environment, a
tool should not force all users into it.

In this paper, we have described our approach to building the MDO
using a patternised scaffold based around existing knowledge
sources. While the work described in this paper allows us to integrate
structured data into an ontology, we are now investigating new ways of
integrating unstructured literate-based knowledge into our ontology;
while we have started the process of formalising, this new knowledge
is far from finished. As described in this paper, though, a
pattern-first, scaffolded approach to ontology development has enabled
us to make significant advances with the MDO. We believe that this
approach is likely to be applicable to many other domains also.

\bibliographystyle{natbib}

\bibliography{2015_scaffolding_pwl_jw.bib}

\end{document}